# A Collaborative Filtering-Based Two Stage Model with Item Dependency for Course Recommendation

Eric L. Lee
Computer Science and
Information Engineering
National Taiwan University
Taipei, Taiwan
r01922164@csie.ntu.edu.tw

Tsung-Ting Kuo
Health Department of
Biomedical Informatics
University of California San Diego
La Jolla, USA
tskuo@ucsd.edu

Shou-De Lin
Computer Science and
Information Engineering
National Taiwan University
Taipei, Taiwan
sdlin@csie.ntu.edu.tw

*Abstract*—Recommender systems have been studied for decades with numerous promising models been proposed. Among them, Collaborative Filtering (CF) models are arguably the most successful one due to its high accuracy in recommendation and elimination of privacy-concerned personal meta-data from training. This paper extends the usage of CF-based model to the task of course recommendation. We point out several challenges in applying the existing CF-models to build a course recommendation engine, including the lack of rating and meta-data, the imbalance of course registration distribution, and the demand of course dependency modeling. We then propose several ideas to address these challenges. Eventually, we combine a two-stage CF model regularized by course dependency with a graph-based recommender based on course-transition network, to achieve AUC as high as 0.97 with a real-world dataset.

## I. INTRODUCTION

Collaborative Filtering (CF) based techniques have become very popular for designing a recommendation system. Among them, the Matrix Factorization (MF) model that jointly learns the user and item latent factors for CF is tremendously successful, and has become one of the standard solutions. This paper extends the existing matrix factorization model to handle a different type of task: recommending courses to the students.

Education institutions normally offer a spectrum of courses in different areas for students to choose, and in many cases the number can be overwhelming. For example, in the year of 2012 there were more than ten thousand courses offered in National Taiwan University (NTU). Thus, selecting suitable courses to take in the upcoming semester is a demanding task for students; more importantly, improper course selection could lead to serious waste of efforts for students and extra administrative burden for faculties.

There have been a few previous works proposed for course recommendation [1, 2]. These models mainly rely on certain meta-information such as the curriculum information from each department, the feedbacks from students, or the grades received by former students. Here we argue that the existence of such meta-data is not guaranteed, and even if it does, those data might not be available due to privacy concern.

To address such concern, our goal is to design a general-purpose, privacy-preserving course recommendation system that requires only minimum personal information (i.e. course registration records) from students. To develop a course recommendation system, first thought would be to treat students as users and courses as items and deploy a CF-based solution. To achieve such a goal, normally we would require some 'ratings' from students to courses, specifying how much they like the course. Then based on the rating, a CF-based solution can utilize the similarity between students and the similarity between courses to predict the level of the students' interests to the courses they have not yet taken.

However, here we argue that there are several practical challenges that hinder the effectiveness of conventional CF models for the course recommendation task:

*(1) Potentially lack of rating data from students to courses.* Traditional CF-based methods rely on the ratings from users to items. However, such rating data might not be universally available for training. For instance, although the partial students' feedback ratings for each course at National Taiwan University exist, they are kept private for privacy concern with only the statistic aggregation of the ratings are made available to the corresponding instructors. On the other hand, it is much less controversial to obtain the information of 'course registration', namely a binary value indicating whether a student registered for a course. Acknowledging the lack of meta-information, in this paper we model the course recommendation task as a One Class Collaborative Filtering (OCCF) challenge for which only partially observable binary matrix indicating whether a student registered for a course is available. For traditional CF problem, each element is either *?* (indicating "unknown") or a score, while for OCCF problem each element is either *?* or *1* (indicating "positive"). Comparing to the traditional CF problem, OCCF encounters more challenges which are detailed in Section 2 and 3.

*(2) Imbalance of the user-item matrix.* The available registration information normally includes the records of current and graduated students. Note that for the current students, only the records of previously registered courses are available. That says, the more senior a student is, the more data are available. We would like to start from performing some analysis on the data to be used to train our model. First, we have realized that normally courses are not taken uniformly across different seniority. Some courses are usually taken earlier, while others are taken in the later stage of a student's academic career. For example, based on our NTU course dataset (during year 2008-2013), Chinese Literature and Calculus are usually taken by freshmen, while the Database Systems are taken by upperclassmen. We found such imbalance of distribution very common. In fact, during 2008~2013, the registration of more than 70% of the courses in NTU are dominated by students of the same level, meaning that more than 50% of the registered students are of the same seniority. Such imbalance of distribution tells us that the course registration records for students of different seniority are likely to be very different. Such difference can cause serious problem for a CF-based model. Assuming now the goal is to recommend courses to a student $s$ who is entering his senior year. A conventional CF model would recommend courses taken by other students that are *similar* to $s$, while the similarity is determined by the number of common courses the two students have taken together previously. However, the imbalance of course-registration record reveals that two students of different seniority (e.g. freshman vs. junior) will not be too similar, and it is very likely the *similar students* to $s$ are of the same seniority. On the other hand, most of the course records from students of the same seniority would not be very helpful for recommendation, since it is likely very few students of the same seniority have taken those courses in the past. It then brings up a dilemma that in order to generate effective recommendation for $s$, a model should indeed include the records of senior or graduated students that have taken higher-level courses. However, based on the definition of CF, those senior people are not necessarily the ones that are the most similar to $s$. Furthermore, the popularity of courses can change over time, thus simply looking up the courses that have been taken by more senior students might not be an ideal solution. This becomes the main challenge this paper tries to handle.

*(3) Courses are coherent and not independent.* Different from product recommendation in which most of the products are independent and the order of purchasing matters little, there are strong temporal and order correlation between courses taken by a student. For example, we would not recommend *calculus* to those who have already taken *advanced calculus*; while recommending *advanced calculus* to those who just took *calculus* seems to be a reasonable idea. Although each department has prerequisites for the courses, this information is usually represented in various unstructured formats (e.g., texts, flowcharts or tables), and thus can hardly be parsed and collected automatically. Therefore, we focus on designing a data-driven method to identify the dependency from data. Conventional CF does not consider such dependency, which becomes another main focus in our solution.

To address the abovementioned challenges, we first adopt the Bayesian Personal Ranking Matrix Factorization (BPR-MF) [3] method, which models the OCCF problem as a ranking problem. Next, we propose a novel two-stage framework to handle the second challenge. In the first stage, we use the registration record of all students to learn the latent features for the courses. Then based on the learned course latent features, in the second stage, we try to learn the student's latent factor to optimize the ranking of courses given a student.

To model the course dependency mentioned in the third challenge, we build an item transition network to capture the probability of a course being taken following another. Such transition network serves two purposes. First, it is used to regularize our two stage CF model; and second, it is used to build a Personalized PageRank model for recommendation. Final results from these two models are then combined to generate our final outputs. We compare our model to several baseline solutions on a real-world course registration dataset which contains about 14K students and about 900K course records. The experiment result shows that the proposed model produces significantly better results, with the final ensemble model reaching 0.97 in AUC.

## II. RELATED WORKS

### A. Course Recommendation Systems

There are several existing works on course recommendation. Parameswaran et al. [2] propose a model using knowledge obtained from curriculum of each department to recommend courses. Bendikir et al. propose a model RARE [1] to discover rules from historical data. However, these solutions require extra meta-information such as the departments the students belong to, the course-registration constraints implemented by each department, and the feedback of students toward each course, which is very different from the goal of not requiring personal meta-information we have setup in this paper.

### B. Collaborative Filtering

Collaborative Filtering (CF) techniques have long been proposed to model explicit feedbacks from users. Comparing to the content-based techniques [4, 5], CF methods are more general, require less information, and in many situations produce superior results. One of the most straightforward CF model is k-nearest neighbor based CF (kNN-CF) [6-8], which is based on user-wise or item-wise similarity for mapping. In general, the similarity measurement (e.g. the Pearson correlation) of kNN-CF is chosen through a trial-and-error process on the validation datasets. Recently, Matrix Factorization (MF) based methods have become popular and are widely accepted as the state-of-the-art single model for CF, as researchers have found that given sufficient rating data, MF methods outperform many other methods in competitions such as KDD Cup [9-13]. Comparing to the kNN-CF methods, MF methods are usually more efficient and effective, as they can discover the latent features which are usually hidden behind the interactions between users and items. However, MF tends to over train data so there has been extensions to address this issue [14, 15]. However, the abovementioned CF techniques are designed to model *explicit* feedbacks from users. In many practical scenarios such as course recommendation, only *implicit* feedbacks are available.[16] Therefore, these CF algorithms cannot be applied directly to deal with tasks such as the recommendation of courses.

## C. One-Class Collaborative Filtering

The problem of applying CF using implicit feedback is known as One-Class Collaborative Filtering (OCCF) task [14]. In OCCF, magnitude of user's preference is usually subtle; therefore, it is hard to distinguish negative examples from unlabeled examples. OCCF can be regarded as a ranking problem in which we need to rank the positive instances above others. Exploiting the idea of ranking optimization, Rendle et al. [3] propose the Bayesian Personalized Ranking Matrix Factorization (BPR-MF) framework to optimize Area Under receiver-operating-characteristic Curve (AUC) for Matrix Factorization (MF) model. In our work, BPR-MF is adopted due to its simplicity and flexibility.

## D. Sequence Recommendation

Sequence recommendation is probably the closest recommendation task to course recommendation, which aims at recommending items to users in the next period. Several studies of sequence recommendation for various applications have been proposed [17, 18]. Although these studies all consider the sequential behavior for recommendation, their methods mostly include ad-hoc scoring functions, special features, or additional information for each specific application. In our paper, we focus on proposing a general method to solve the real-world course recommendation problem without using meta-data or application-specific content information (e.g., detailed student or course information). The CF-based solution proposed by Rendle et al. [19] for sequence recommendation is relevant to our problem, but it cannot be directly applied to deal with course recommendation. First, it assumes sparse transition behavior and longer period in training. In our problem, the transition behavior (course registration) is dense, and the number of temporal slot is few. Thus, it is hard to learn a meaningful MC. Also, the issue of distribution imbalance is not handled in this model.

## III. METHODOLOGY

We are given a registration matrix $R$ indicating whether a student $s$ registered a course $c$ in the past. That is, in matrix $R$ an element is $1$ if the student did take the course, and $?$ (indicating "unknown") if the student has not yet registered for the course. We model the course recommendation problem as: given the Registration Matrix $R$, predicting the probabilities of the *unknown* elements actually being $1$. To solve this problem, we propose a two-stage CF model with dependency regularization. In the following subsections, we introduce the three main components of our model: Bayesian Personal Ranking Matrix Factorization (BPR-MF), two-stage training, and course dependency regularization.

### A. Bayesian Personal Ranking Matrix Factorization (BPR-MF)

We first introduce a variation of the well-known Matrix Factorization (MF) technique to be adopted into our scenario. Basic MF model aims at finding two matrices, $P$ (the latent feature matrix for students) and $Q$ (the latent feature matrix for courses), of which their multiplication can best recover the input matrix $R$ minimizing the square error between $R_{sc}$ and $P_s \cdot Q_c$. The objective function of MF is,

$$\min_{P,Q} \sum_{(s,c) \in W} (R_{sc} - P_s \cdot Q_c)^2 , \qquad (1)$$

where $W$ represents the set of all existing (*student*, *course*) pairs. Eventually MF learns the values in the $P_s$ and $Q_c$ vector as the latent student/course features, and uses the corresponding inner product of $P$ and $Q$ to produce the prediction score. However, applying MF technique directly to solve our problem leads to a serious drawback. In the given scenario, the values in the registration matrix $R$ are either one (i.e., registered) or unknown. Directly applying MF with such data will lead to a useless solution where the model predicts every entry as $1$ to minimize the error.

The problem we are trying to solve is known as the One Class Collaborative Filtering (OCCF). To solve OCCF, we adopt the Bayesian Personal Ranking Matrix Factorization (BPR-MF) [3] model. BPR-MF optimizes the Area Under receiver-operating-characteristic Curve (AUC) instead of square error. The reason to optimize AUC is that it yields a model that produce faithful ranking of instances, giving higher prediction score to the courses to be recommended.

Given a student $s$, denote the courses that $s$ has taken as a set $I_s^+$ and those $s$ has not taken as $I_s^-$. BPR-MF maximizes the difference between the likelihood that the courses are in $I_s^+$ and the likelihood that the courses are in $I_s^-$. Let $P_{sk}$ be the *k-th* latent feature for student $s$ in $P$, $Q_{ik}$ be the *k-th* latent feature of a taken course $i \in I_s^+$ in $Q$, and $Q_{jk}$ be the *k-th* latent feature of an untaken course $j \in I_s^-$ in $Q$, the objective function is defined as:

$$Error := \sum_{(s,i,j) \in D} \ln(1 + \exp(-y_{sij})) + \lambda(|P|^2 + |Q|^2) , \qquad (2)$$

where

$$D := \{(s, i, j) | i \in I_s^+, j \in I_s^-, and\ s \in S\} \qquad (3)$$

is the tuple of positive-negative courses for the student $s$, and

$$y_{sij} = \sum_k^K P_{sk}(Q_{ik} - Q_{jk}) \qquad (4)$$

is the pairwise difference of the likelihoods between $I_s^+$ and $I_s^-$, $K$ is the number of latent features, and $\lambda$ is a regularization parameter. We can further define

$$u_{sij} = \frac{-\exp(-y_{sij})}{1+\exp(-y_{sii})}. \quad (5)$$

Then, BPR-MF can be learned in Stochastic Gradient Descent (SGD), after deriving the partial derivatives of $P_{sk}$, $Q_{ik}$, and $Q_{jk}$ as:

$$\frac{\partial Error}{\partial P_{sk}} = u_{sij} \cdot (Q_{ik} - Q_{jk}) + 2\lambda P_{sk},$$

$$\frac{\partial Error}{\partial Q_{ik}} = u_{sij} \cdot (P_{sk}) + 2\lambda Q_{ik},$$

$$\frac{\partial Error}{\partial Q_{jk}} = u_{sij} \cdot (-P_{sk}) + 2\lambda Q_{jk}. \quad (6)$$

However, the complexity of SGD-based optimization procedure requires $O(|I_s^+| \cdot |I_s^-|)$. This can cause significant computational burden because $|I_s^-|$ is usually very large (i.e., there are many courses a student has not yet taken). To overcome such limitation in the course recommendation problem, we perform down-sampling on the negative set $I_s^-$. That is, for a given student $s$, we randomly sample one negative course in $I_s^-$ for every instance in $I_s^+$. This way, the training speed is significantly improved.

*B. Two-Stage Training for BPR-MF*

Although the BPR-MF model is a nice solution for the OCCF problem, we still need to handle the issue of distribution imbalance of taken courses as described previously (challenge 2 in Section 1). To elaborate the issue and our solution, we first divide the course registration records into four parts (as shown in Figure 1). First, the students are divided into two parts: the *graduated students* whose course selection information throughout their academic career are available; and the *current students* who require the course recommendation service for the upcoming semester. For the current students, we can only obtain their registration data from previous years. Then, we divide the courses roughly into two parts: the *fundamental courses* that are more likely to be taken by lowerclassmen, and the *advanced courses* that are more likely to be taken by upperclassmen.

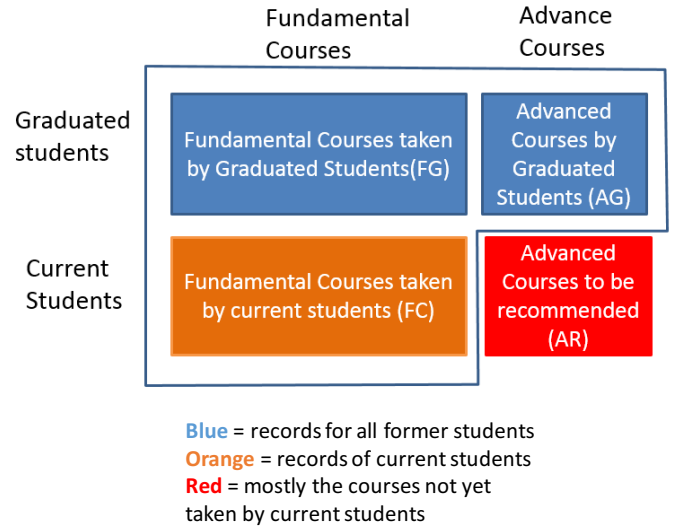

Fig. 1. Four parts of the course registration records.

The matrix can therefore be divided into four parts: (1) The upper-left part represents the *fundamental courses* that have been taken by the *graduated students* (*FG*); (2) The upper-right part represents the *advanced courses* that have been taken by the *graduated students* (*AG*); (3) The lower-left part represents the *fundamental courses* that have been taken by the *current students* (*FC*); and (4) The lower-right part represents the *advanced courses* that are more likely to be taken by the *current students* in the upcoming year (*AR*). Note that *AR* is supposed to be much sparser than *AG*, since not as many *current students* have taken *advanced courses*. As result, the BPR-MF model tend to downgrade the probability inside *AR*, which leads to an inferior model that tends to *NOT* recommend *advanced courses* to the *current students*, which is opposite to our goal.

Our idea to address such deficiency is to learn the course latent features and student latent features separately in two stages. The latent course features in fact can be used to represent the similarity between courses, namely whether two courses are taken by a similar set of students. To take advantage of such latent features of courses that preserve the connection between the *fundamental courses* and *advanced courses*, we propose to train a BPR-MF model using data from *FG*, *AG*, and *FC*, excluding *AR*. Also during the negative sampling stage, entries in *AR* cannot be sampled as negative to avoid bias. The latent courses features are kept as the input to the second stage. In the second stage, we fix $Q$ and perform the BPR-MF model again only using data from *FC*. The goal is to learn refined latent features for only *current students*. Since this time $Q$ is fixed, meaning that the learning algorithm should respect the dependency learnt between *fundamental courses* and *advanced course* while training the student latent features. Note that there is no need to use *FG* to learn the latent features of *graduated students* since they do not need recommendation anymore. The final prediction is obtained from the inner product of the fixed $Q$ and newly obtained $P'$ in the second stage. The details are shown in Algorithm 1.

It should be noted that the latent student features in the decomposed student matrix $P$ cannot be utilized directly. As abovementioned, the latent student features for *current students* tend to have lower preference towards *advanced courses*.

### C. Item Transition Network for Regularizing Two-Stage BPR-MF

By applying the two-stage training for BPR-MF, we can now address the issue of data imbalance. However, we argue that the explicit *dependency* between courses should be modeled as well. That is, certain courses are more likely to be taken right after another. Here we first propose the item transition network to model the dependency between courses. Next, such network is exploited as a regularization term in our two-stage BPR-MF model. The goal is to strengthen the connection between dependent courses.

Algorithm 1. Two-Stage BPR-MF.

---

**Input**: $FG$, $AG$, $FC$, $\alpha$

**Output**: $P \cdot Q$

// First Stage: train $Q$ using $FG$, $AG$, $FC$

1: **repeat**

2:    **for** $(u, i) \in \{FG, AG, FC\}$ **do**

3:      Draw a negative sample $j$ from $I_s^-$ in $\{FG, AG, FC\}$

$$P_{sk} := P_{sk} - \alpha \cdot \frac{\partial Error}{\partial P_{sk}}$$

$$Q_{ik} := Q_{ik} - \alpha \cdot \frac{\partial Error}{\partial Q_{ik}}$$

$$Q_{jk} := Q_{jk} - \alpha \cdot \frac{\partial Error}{\partial Q_{jk}}$$

4:    **end for**

5: **until** convergence

// Second Stage: fix $Q$, and then update $P$ using $FC$

6: **repeat**

7:    **for** $(u, i) \in FC$ **do**

8:      Draw a negative sample $j$ from $I_s^-$ in $FC$

$$P_{sk} := P_{sk} - \alpha \cdot \frac{\partial Error}{\partial P_{sk}}$$

9:    **end for**

10: **until** convergence

---

*1) Constructing the Item Transition Network:* We propose a directed, weighted, homogeneous graph called the item transition network, to model the dependency between the courses. Intuitively, an item transition network is defined as follows: (1) Node: a course; (2) Link: the dependency between two courses; the source node represents the course taken in a year, and the destination node represents the course taken in the next year; and (3) Weight of link: ranges from 0 to 1, representing the weighted probability that the target course is taken in the next year after the source course is taken. We illustrate the construction of item transition network using an example (Figure 2 (top)). In this example, there are three students ($s_1$, $s_2$, and $s_3$), four courses (A, B, C and D), and three grade levels. The three students are all now in their junior year. Student $s_1$ took course A and B in his freshman year, C in the sophomore year, and D in the junior year. The course records for student $s_2$ and $s_3$ are also represented in the similar way in Figure 2 (top). The item transition network constructed from this example of course records is shown in Figure 2 (bottom). We demonstrate the generation of the weights using out-links of node A (i.e., the probabilities that students take course A in a year, and then take course B, C or D in the next year). There is exactly one student ($s_3$) who takes course B the year after he/she takes course A. There are two students ($s_1$, $s_2$) who take course C the year after they take course A. There is only one student ($s_2$) who takes course D the year after he/she takes course A. Therefore, the weight of the link from A to B is 1 / (1+2+1) = 0.25, the weight of the link from A to C is 2 / (1+2+1) = 0.5, and similarly the weight of the link from A to D is 1 / (1+2+1) = 0.25. In this way, we can compute the weights for all links in the network, as shown in Figure 2 (bottom). Note that only the links with non-zero weights are included.

| Student | Freshmen | Sophomore | Junior |
|---|---|---|---|
| $s_1$ | A B | C | D |
| $s_2$ | A | C D | B |
| $s_3$ | C | A D | B |

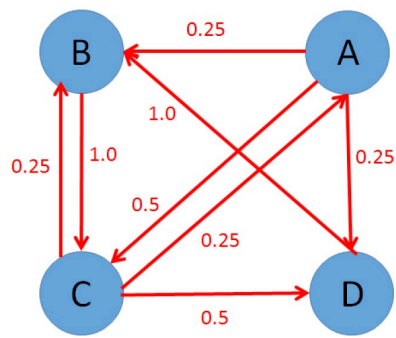

Fig. 2. (top) Example course records for three students, four courses, and three grade levels. (bottom) The item transition network for this example.

*2) Regularizing the Two-Stage BPR-MF using Item Transition Network:* To incorporate the course dependency

*information modeled in the item transition network, we add a soft constraint on the connected courses in our two-stage BPR-MF model. Inspired by Ma et al. [20], we propose to impose such soft constraint by adding a regularization term into our model. That is, we add the Course Dependency Regularization (CDR) term to the original BPR-MF error function (2) as below:*

$$Error' := Error + \frac{\beta}{2} \sum_{f \in I} \sum_{g \in N(f)} w(f,g) \cdot |Q_f - Q_g|^2 , \quad (7)$$

where $\beta$ is a positive constant parameter, $I$ is the set of all courses, $N(f)$ is the set of courses in the item network pointed to by course $f$, and $w(f, g)$ is the weight of the link from $f$ to $g$.

And the partial derivatives are as follows:

$$\frac{\partial Error'}{\partial P_{sk}} = \frac{\partial Error}{\partial P_{sk}}, \frac{\partial Error'}{\partial Q_{ik}} = \frac{\partial Error}{\partial Q_{ik}} + t_f, \frac{\partial Error'}{\partial Q_{jk}} = \frac{\partial Error}{\partial Q_{jk}} + t_f . \quad (8)$$

where

$$t_f = \beta \sum_{g \in N(f)} w(f,g) \cdot (Q_f - Q_g) . \quad (9)$$

We then apply the above partial derivatives to Algorithm 1, as the two-stage BPR-MF model with CDR. Furthermore, we have observed from data that the item transition network is dense (average node degree is 99.8), which can seriously hurt the training efficiency. In practice, we simply apply a weight threshold $T$ to remove edges whose weight is small, which leads to a much sparser network for training.

### D. Personalized PageRank (PPR) and Ensemble

Besides using the information in the item transition network as a regularization term, another way to utilize the course dependencies is to perform centrality algorithm (such as random walk) on the item transition network to identify important courses for recommendation based on dependencies. That is, we can rank and recommend the nodes (courses) based on their importance in the item transition network. Therefore, we design a Personalized PageRank (PPR) algorithm to recommend courses to each student. The main differences from the original PageRank algorithm with a damping factor $\gamma$ are as follows:

(a) Because the courses we would like to recommend is for next period, we limited the start/restart nodes to a set $C$, which includes only the courses the student has taken at the current time. For example, in our experiment we would like to recommend courses to senior students, so in the PPR algorithm we limit the start/restart nodes to the set of courses that are taken in the junior level.

(b) With probability $\gamma$, the algorithm walks to a neighbor with probability proportional to the weight of the link to that neighbor.

(c) With probability $1 - \gamma$, the algorithm restarts from a randomly selected node in a subset $C$ of the nodes of the item transition network.

Finally, as shown in several previous studies [11], an ensemble of CF-based models and graph-based models can significantly improve the performance of a recommendation system. Here we propose to combine the proposed two-stage BPR-MF model with dependency regularization and the PPR as our final model for recommendation. Since both models produce the ranking results, we ensemble them using linear RankSVM [21], a supervised ranking-based model.

### IV. EXPERIMENTS

To evaluate the proposed ideas, we compare our model with several algorithms using real-world course registration records. The dataset, experiment settings, comparing methods, results, discussion, and statistical tests are described in the following subsections.

### A. Dataset and Experiment Settings

We collect 6 years of course registration records for all NTU students from 2008 to 2013. That says, for those started as a freshman in NTU at 2008, 2009, and 2010, we have full 4-year registration records for them. For evaluation purpose, we ignore the students whose 4-year registration records are incomplete. The dataset contains 13,977 students (class 2008 = 4,736, 2009 = 4,686, and 2010 = 4,555) and 896,616 course registration records (class 2008 = 311,283, 2009 = 299,772, and 2010 = 285,561).

In our experiment, we aim at recommending the advanced courses to students who are entering their senior year. We believe it is a more useful course recommendation scenario since freshman and sophomore students generally need to take required entry level courses, and therefore has less freedom in course selection. On the other hand, senior students have much higher flexibility to choose courses of their interests. That says, we have the ground truth of senior course registration data for class 2008, 2009, and 2010 students. We use all the data except the senior registration record of class 2010 students, which is used as testing data, as the training data. From the training set, we further choose the senior year registration records of class 2009 as the validation set to tune the parameters of all the competing models.

The parameters we used in our experiment are as below: $K$ (the number of latent features in BPR-MF) = 12; $\lambda$ (the regularization parameter in BPR-MF) = 0.05; $\alpha$ (the learning rate in BPR-MF) = 0.05; $\beta$ (the regularization parameter using Item Transition Network) = 0.008; $T$ (the weight threshold for Item Transition Network) = 0.03; and $\gamma$ (the damping factor in PPR) = 0.7. We choose the Area Under receiver-operating-characteristic Curve (AUC) as the evaluation metric. For each student, we rank the predicted scores for all courses, and compare with gold standard registration records from year 2013 in the test set, to calculate the AUC of each student. Then, we report average AUC of all students.

*B. Compared Methods*

We compare four different sets of models in the experiment: memory-based models, graph-based models, our BPR-MF solutions, and ensemble of different types of solutions.

One potentially powerful baseline is to always recommend the most popular courses. Thus, we also implement a simple non-personalized baseline to recommend the users the most popular courses based on historical data. For memory-based methods, the idea is to recommend courses base on the similarity of students. The courses taken by similar peers have higher chances to be recommended. Therefore, we first calculate the similarity of student pairs based on their course registration data. Then, we can calculate the score of a student $s$ to a course $c$ according to this similarity: $score(s,c) := \sum_{s' \in S \cap c} sim(s',s)$, where $S$ is the set of all senior or graduated students, $s' \in S \cap c$ represents the students that have taken course $c$, and $sim(s', s)$ is the similarity of two students s' and s. Note that in the similarity-based model we focus on finding the similarity between a student and his/her senior peers, rather than students of the same grade, since experiment shows that bringing the students of the same level into consideration can hurt the performance. It is reasonable as such records might carry negative bias toward advanced courses. We apply the following two similarity functions in the memory-based CF models (1) number of intersectional courses, and (2) Jaccard similarity.

TABLE I.  EXPERIMENT RESULTS

| Category | Method | AUC |
|---|---|---|
| Baseline | Course Popularity | 0.8172 |
| Memory-based | Number of Intersectional Courses | 0.8678 |
| | Jaccard Similarity | 0.8922 |
| Graph-based | Personalized PageRank | 0.9334 |
| Model-based | BPR-MF | 0.9366 |
| | BPR-MF + Two-Stage Training | 0.9404 |
| | Our Model (BPR-MF + Two-Stage Training + Course Dependency Regularization) | 0.9427 |
| Ensemble | Our Model + Personalized PageRank | 0.9709 |

*C. Results and Discussion*

The experiment results are shown in Table 1. It shows the baseline popularity-based model performs the worst, which is reasonable as it is not personalized. Memory-based models perform better than the baseline, but not as promising as the other models, which is consistent with the outcomes of the conventional recommendation task. The graph-based model using item transitional network performs better than the memory-based models, meaning that modeling the item transition is more useful than modeling the user similarity in course recommendation. The BPR-MF model performs slightly better than the graph-based model, probably because in BPR-MF both the item similarity and user similarity are considered. Adding two-stage training into BPR produces significant improvement (see next section for the hypothesis tests), and adding CDR can further boost the performance. Interestingly, we have realized that combining the graph-based solution with our model creates a large jump on the performance, reaching 0.9709 from 0.9427.

*D. Hypothesis Tests*

Although Table 1 shows improvements of using two-stage training and CDR, we would like to perform deeper analysis to evaluate the effectiveness of these two methods. Thus, we conduct the following two hypothesis tests: (a) Test 1: comparing *BPR-MF + Two-Stage Training* (target model 1) with *BPR-MF only* (original model 1). (b) Test 2: comparing *BPR-MF + Two-Stage Training + CDR* (target model 2) with *BPR-MF + Two-Stage Training* (original model 2). For both tests, we first calculate the difference of the AUC between two models (i.e., AUC of the target model minus AUC of the original model) for each of the 4,555 students in the test set (class 2010), and then perform a hypothesis test. We set the significance level to 0.05 and generate the P-value as a measure to accept or reject the hypotheses. The results show that the P-value for Test 1 is $3.24 \times 10^{-12}$. It shows that the two-stage model is significantly better than the original BPR-MF. On the other hand, the P-value for Test 2 is 0.0135, also demonstrates the usefulness of CDR.

V. CONCLUSIONS

The capability to recommend items in real-world setting is highly valuable in practice, as in real world the ratings may only be binary, the training data for a new period may be missing, or the relationship among the items to be recommended may be complicated. These issues are common in application domains such as the course recommendation task. In this paper, we demonstrate how such a challenging recommendation task can be solved using a two-stage collaborative filtering model with dependency regularization. We show how the one-class issue can be mitigated using BPR-MF method, propose a novel two-stage training method to learn the parameters using incomplete training data, and devise a transition network to integrate the item dependency as the regularization term in our model. Most importantly, with growing awareness of privacy, our method provides a way for applications that tries to recommend items with no content information. It should be noted that our current model is mainly focused on recommending courses for upperclassmen (e.g. students in junior or senior year) and might not be very effective for lowerclassmen due to lack of registered courses for training. The future work will be focused on extending the model to deal with cold start students.

ACKNOWLEDGMENT

This material is based upon work supported by the Air Force Office of Scientific Research, Asian Office of Aerospace Research and Development (AOARD) under award number No.FA2386-17-1-4038.